\documentstyle[12pt]{article}
\newcommand{\bal}{\bar \alpha}
\newcommand{\bef}{$\beta$-function}
\newcommand\mysection{\setcounter{equation}{0}\section}
\def \as{\relax\ifmmode\alpha_s\else{$\alpha_s${ }}\fi}
\def\MSbar{\relax\ifmmode\overline{\rm MS}\else{$\overline{\rm MS}${ }}\fi}
\def\PT{{\mbox{\scriptsize PT}}}

\def\PH{{\mbox{\scriptsize PH}}}
\def\cC{{\cal C}}
\def\e{{\mathop{\rm e}}}
\def\L{\Lambda_{\mbox{\scriptsize QCD}}}

\def\baeq{\begin{appeq}}     \def\eaeq{\end{appeq}}
\def\baeeq{\begin{appeeq}}   \def\eaeeq{\end{appeeq}}
\newenvironment{appeq}{\beq}{\eeq}
\newenvironment{appeeq}{\beeq}{\eeeq}

\renewcommand{\theequation}{\thesection.\arabic{equation}}

\newcounter{hran}
\renewcommand{\thehran}{\thesection.\arabic{hran}}

\def\bmini{\setcounter{hran}{\value{equation}}
\refstepcounter{hran}\setcounter{equation}{0}
\renewcommand{\theequation}{\thehran\alph{equation}}\begin{eqnarray}}

\def\bminiG#1{\setcounter{hran}{\value{equation}}
\refstepcounter{hran}\setcounter{equation}{-1}
\renewcommand{\theequation}{\thehran\alph{equation}}
\refstepcounter{equation}\label{#1}\begin{eqnarray}}

\def\emini{\end{eqnarray}\relax\setcounter{equation}{\value{hran}}\renewcommand{\theequation}{\thesection.\arabic{equation}}}

\def\cO#1{{\cal O}\left(#1\right)}

\def\ga{\mathrel{\mathpalette\fun >}}
\def\fun#1#2{\lower3.6pt\vbox{\baselineskip0pt\lineskip.9pt
  \ialign{$\mathsurround=0pt#1\hfil##\hfil$\crcr#2\crcr\sim\crcr}}}

\def\eV{{\rm e\kern-0.12em V}}            
  
\def\half{{\textstyle {1\over2}}}

\def \al {\relax\ifmmode{\alpha}\else{$\alpha${ }}\fi}
\def \be {\relax\ifmmode{\beta}\else{$\beta${ }}\fi}
\def\ga{\gamma}
    
\def\Im{\mathop{\rm Im}}    \def\Re{\mathop{\rm Re}}

\def\abs#1{\left| #1\right|}

\def\ben{\begin{enumerate}}  \def\een{\end{enumerate}}
\def\bit{\begin{itemize}}    \def\eit{\end{itemize}}
\def\beq{\begin{equation}}   \def\eeq{\end{equation}}
\def\beeq{\begin{eqnarray}}  \def\eeeq{\end{eqnarray}}
\def\bq{\begin{quote}}       \def\eq{\end{quote}}

\newskip\humongous \humongous=0pt plus 1000pt minus 1000pt
\def\caja{\mathsurround=0pt}
\def\eqalign#1{\,\vcenter{\openup1\jot
\caja   \ialign{\strut \hfil$\displaystyle{##}$&$
\displaystyle{{}##}$\hfil\crcr#1\crcr}}\,}

\newif\ifdtup

\def\eqal2#1{\,\vcenter{\openup1\jot
\caja   \ialign{\strut \hfil$\displaystyle{##}$&\hfil$
\displaystyle{{}##}$\hfil &$
\displaystyle{{}##}$\hfil\crcr#1\crcr}}\,}

  \def\refup#1{~$^{\cite{#1}}$}

 \def\cite#1{[\ref{#1}]}

\begin{document}

\begin{titlepage}
\renewcommand{\thefootnote}{\fnsymbol{footnote}}

\begin{flushright}
CERN-TH/95-328\\
hep-ph/9512407
\end{flushright}
\vspace{.3cm}
\begin{center} \LARGE
{\bf Are IR Renormalons a Good Probe \\
for the Strong Interaction Domain?}
\end{center}
\vspace*{.3cm}
\begin{center} {\Large
Yu.L. Dokshitzer and N.G. Uraltsev\\
\vspace{.4cm}
{\normalsize
{\it TH Division, CERN, CH-1211 Geneva 23,
Switzerland}\\
and\\
{\it St.Petersburg Nuclear Physics Institute,
Gatchina, St.Petersburg 188350, Russia}\footnote{Permanent address}\\
\vspace{.3cm}
\vspace*{.4cm}
}}

{\vspace*{.2cm}\Large{\bf Abstract}\\}
\end{center}
\begin{quote}
We study the origin of non-analyticity in $\as$ of a short-distance QCD
observable to demonstrate that the infrared renormalons,
the same-sign factorial growth of the perturbative expansion, is a universal
phenomenon that originates entirely from the {\em small}\/ coupling
domain.
In particular, both the position and the nature of the singularity of the
Borel transform of the perturbative series prove to be independent of
whether the running coupling $\al(k^2)$ becomes singular at some
finite scale (``Landau pole''), or stays finite down to $k^2\!=\!0$.
We argue that getting hold of the infrared renormalons {\em per se}\/
can help next to nothing in quantifying non-perturbative effects.
\end{quote}
\vspace*{\fill}
CERN-TH/95-328\\
December 1995
\end{titlepage}
\addtocounter{footnote}{-1}

\newpage

\mysection{Introduction}

Last years showed revival of interest to the problem of asymptotic
behaviour of the perturbative (PT) expansion in QCD \cite{recent}.
Considering the Borel transform of PT series
has been suggested \cite{ren}
as technical means of improving convergence, and
demonstrated that the PT expansion by itself prompts about nontrivial
dynamics at low momentum scales via infrared (IR) renormalons.

In reality, one does not expect\refup{ren} QCD observables to have proper
analyticity in $\as$ in the vicinity of $\as\!=\!0$, which is necessary
to justify
the Borel resummation as a mathematically well defined operation.
The original arguments rely on the nontrivial analytic properties
of QCD observables with respect to the external momentum scale
parameter $Q^2$. These properties are driven by causality and unitarity
constraints, supplemented with the pattern of the hadron spectrum.
On the other hand, the IR renormalons themselves have a simpler origin
related merely to the nature of the perturbative expansion.
In the present paper we address these features of the PT expansion
to illustrate that the IR renormalons are not a reflection
of the actual non-perturbative (NP) dynamics.
Rather they are an artefact of an attempt to describe physics
occurring at quite different scales by means of one and the same expansion
parameter.
The salient conclusions we draw from our analysis are:

$\bullet$
IR renormalons persist even if the coupling stays finite
at arbitrarily small momentum scales when no apparent uncertainty
associated with the Landau singularity shows up.

$\bullet$
The position and the nature of the IR renormalon singularity in the Borel
image of a generic hard QCD observable are determined by the first
two coefficients of the \bef.

$\bullet$
The modification of the Borel integration prescription one can design to
represent the true answer, depends heavily on the details of strong interaction
dynamics as well as on the particular observable.
Therefore, such a possibility seems to carry no practical value
as a perturbative algorithm.

$\bullet$
The $Q^2$-dependence of a short distance observable, inferred from the Borel
resummation, typically yields an incorrect image of the actual magnitude of
condensate effects, when considered at intermediate $Q^2$.

We also briefly address the OPE-motivated procedure to illustrate
that it is free from the IR renormalon problems.

In this paper we outline the main framework and the results
of the analysis supplying minimal illustrations when necessary.
A more complete discussion with better elaborated qualitative
and quantitative considerations
and deeper physical arguments about the relevance
of the analysis undertaken,
will be presented in a forthcoming publication \cite{DUbig}.

\mysection{Toy model}

To study the origin of the IR renormalons and their relation to NP
dynamics we concentrate on a simplified model peeled off inessential details.
To this end, we address the problem of the PT expansion
of
the integral
\beq\label{observ}
 I(\al) = \int_0^{Q^2} \frac{dk^2}{k^2} \left(\frac{k^2}{Q^2}\right)^p
\al(k^2)\>
\eeq
as a model for a short-distance dominated observable.
Integrals of this type naturally emerge, for example,
when one calculates the one-gluon correction to
a collinear safe QCD observable using the running coupling
in the integrand.
Correctness of this substitution can be unambiguously proven in an
Abelian theory; in the QCD context such a doing is often motivated
by the ``naive non-Abelization'' or by the ``extended BLM prescription''
for fixing the relevant hardness scale\refup{blm} in Feynman integrals.

It is easy to see that the observable (\ref{observ}) satisfies the
inhomogeneous renormalization group (RG) equation
\beq\label{RGeq}
\beta(\alpha) \frac{dI(\alpha)}{d\alpha} \;+\; p\,I(\al)\>=\>\al\>,
\eeq
where
\beq\label{beta}
 \frac{d \al(k^2)}{d\log{k^2}}\>=\> \beta\left(\alpha(k^2)\right).
\eeq
Hereafter, to simplify notation, we absorb the factor $\be_0/4\pi$ into
the coupling and denote by $\al$ (without an argument)
its value at the external hard scale of the problem,
$$
 \al \>\equiv\> \frac{\be_0}{4\pi}\, \as(Q^2)\>.
$$
In this notation the PT expansion of the QCD $\be$-function is
\beq\label{betaexp}
  \be(\al) \>=\> - \al^2 - \gamma \al^3 + \>\mbox{higher terms}\,,
\quad \gamma=\frac{\be_1}{\be_0^2}>0\,.
\eeq
In this Section we concentrate on the properties of $I$ as a function of $\al$
(its perturbative expansion, Borel representation, analyticity in $\al$).
The analytic properties of $I(\al)$ stem from the dependence of $\al(k^2)$
 on $\al$ at $k^2<Q^2$, which dependence is determined by the RG
{\em trajectories}\/
\beq\label{traj}
  \int_{\al}^{\al(k^2)} \frac{d\al'}{-\be(\al')} =\ln \frac{Q^2}{k^2} \>
\equiv\> t\,.
\eeq
In what follows we shall refer to $t$ as ``time''.

It is clear that the integral (\ref{observ}) is sensible only if
$\al(k^2)$ does not develop a Landau singularity at finite positive $k^2$
(at finite time), which implies that $\beta(\al)/\al$
has a zero on the positive real axis.
(The first such zero will be generically denoted by $\bal$,
$0<\bal\le +\infty$.)
Arguments in favour of an ``infrared finite'' coupling
as the only reasonable expansion parameter for QCD observables,
at least in the present context,
will be given in a more detailed publication \cite{DUbig}.

Our aim is to compare the ``exact'' expression (\ref{observ})
with the results one obtains using the Borel resummation tricks.

\subsection{Analyticity in $\al$: perturbative and physical ``phases''}
It is straightforward to solve the RG equation
(\ref{RGeq}).
First,  one finds the solution of the {\em homogeneous}\/ equation
\bminiG{homo}
\label{homoeq}
&& \be(\al) \frac{d\,X(\al)}{d\al}\,+\,pX(\al)\,=\,0\>,\quad
 X(\al)=\left(f(\al)\right)^p\>; \\
\label{homosol}
&& f(\al)= \exp\left\{ -\int^{\al}\frac{d\al'}{\be(\al')}\right\}.
\emini
Function $f(\al)$ is nothing but the RG-invariant expression for a
pure power of the momentum scale. For small $\al$, for example,
one invokes the expansion (\ref{betaexp}) to obtain,
up to an overall factor,
\beq\label{fappr}
 f(\al)=\exp\left\{- \frac{1}{\al}\right\}\al^{-\gamma}\left(1+\cO{\al}\right)
\>=\>\frac{\mbox{const}}{Q^{2}}\,.
\eeq
In terms of $f(\al)$ one can write the solution of the inhomogeneous equation
in the form
\beq\label{Isol}
  I(\al)= f^p(\al)\int_{\al_\infty}^{\al} \frac{d\al'}{\be(\al')}
          \frac{\al'}{f^p(\al')} \,.
\eeq
Here $\al_\infty$ is the asymptotic value of the running coupling
at $t\!=\!\infty$ along the trajectory (\ref{traj}) that starts
from a given $\al$ at $t\!=\!0$.
We shall call such  point(s) {\em attractive}.
In the {\em physical phase}, that is for real positive (small)
initial $\al$ values, $\al_\infty=\bar{\al}\,$:
\beq\label{IPH}
  I^\PH(\al)\>= \>
   \int_{\bal}^{\al} \frac{d\al'\>\al'}{\be(\al')}
\exp\left\{ -p\int^{\al}_{\al'}\frac{d\al''}{\be(\al'')}\right\}.
\eeq
{}From explicit expressions (\ref{Isol}), (\ref{fappr}) it is clear
that an analytic continuation of $I(\al)$, starting from some small
positive $\al$, can fail -- for small $\al$ --
only due to non-analyticity in $\al$
of the lower limit of the integral,  that is $\al_\infty(\al)$.
Let us demonstrate that it does fail; the faster, the smaller initial
value of $\al$ is taken.

To this end let us look at the RG trajectories $\al(t)$ for complex
initial values of $\al$.
We start with the case when $\abs{\al}$ is small but its phase is finite.
In this case the value of $\abs{\al(t)}$ stays uniformly small along the
whole trajectory, $0\!\le\! t\!\le\! \infty$,
so that it suffices to approximate the full \bef\
by its one-loop expression to find
\beq
\frac{1}{\al(t)} = \left(\frac{1}{\al} - t \right)
\left[\,1+\cO{\al(t)\ln\al(t)}\,\right] \>\approx\> \frac{1}{\al} - t\,.
\label{24}
\eeq
In this approximation trajectories lie on small circles
(either in the upper or in the lower half-plane)
that touch the real axis at $\al\!=\!0$.
The radius of a circle is related to the phase of the initial $\al$ by
\beq\label{radius}
r= \frac{1}{2}\abs{\Im {\al}^{-1}}^{-1}=
\frac{1}{2}\frac{|\al^2|}{|\Im \al|} \>\ll\> 1\,.
\label{25}
\eeq
It is important to realize that the existence of the family of such
trajectories having the common limiting point
$\al_\infty\equiv \bal_0 =-0$ is a universal, purely {\em perturbative}\/
feature of the asymptotically free theory with $\be\propto -\al^2$.
In QCD the double-zero of the \bef\ acts as a {\em repulsive}\/ point for
trajectories with $\al\!>\!0$ and as an {\em attractive}\/ one for $\al\!<\!0$.
The domain in the complex $\al$-plane covered by these trajectories will be
called the ``perturbative phase''.
Within this domain $I(\al)$ is an analytic function but explicitly
{\em different}\/ from the physical answer (\ref{IPH}).
So, we define
\beq\label{IPT}
  I^\PT(\al)\>= \>
   \int_{-0}^{\al} \frac{d\al'\>\al'}{\be(\al')}
\exp\left\{ -p\int^{\al}_{\al'}\frac{d\al''}{\be(\al'')}\right\}.
\eeq
With the decrease of the phase of $\al$, the radius (\ref{radius})
increases and trajectories start to distort, being affected by the higher
terms in the \bef.
At some critical point ($r\sim 1$, in general) a bifurcation occurs
and the trajectories switch to another attractive point, different from
$-0$.
Such a jump results in a {\em singularity}\/ (non-analyticity)
of $I(\al)$.
It is easy to see from (\ref{radius}) that this happens for
{\em almost real}\/ initial $\al$ values:
\beq
 \frac{\Im\al}{\Re\al} \>=\> \frac{c\abs{\al}^2}{\Re\al}
\approx c\abs{\al}\approx c\Re\al \>\ll\> 1\,.
\eeq
In the general case, the topological structure of the analyticity domains
and, correspondingly, the number of jumps $I(\al)$ experiences while
approaching the real axis, can be quite complicated, reflecting
the structure of the attractive zeroes of the \bef\ in the complex $\al$-plane.
For the sake of simplicity we shall restrict our wording to the
scenario when the very first bifurcation (``phase transition'' from the
perturbative domain) leads us directly to the physical phase;
the generalization is straightforward.

To illustrate this phenomenon one can consider two simplest examples
with the two attractive points, $-0$ and $\bal$,
with both  examples respecting the first two PT terms of the
QCD \bef\ (\ref{betaexp}).
In the first model with a polynomial $\be$
the physical coupling freezes at a finite value $\bal$:
\bminiG{twoex}
\label{ex1}
 \be(\al) = -\al^2(1+\al/\al_1)(1-\al/\bal)\>, \quad
{\al_1}^{-1}-\bal^{-1} = \gamma\>.
\eeeq
The second model employing a rational \bef\
possesses a pair of complex conjugated poles and yields
$\al(k^2)$ increasing logarithmically with $k^2\!\to\!0$,
that is $\bal\!=\!\infty$:
\beeq
\label{ex2}
\be(\al) = -\frac{\al^2}{ (1-\half\gamma\al)^2 + h^2\al^2 }\>.
\emini

\begin{figure}
\vspace{7.0cm}
\includegraphics{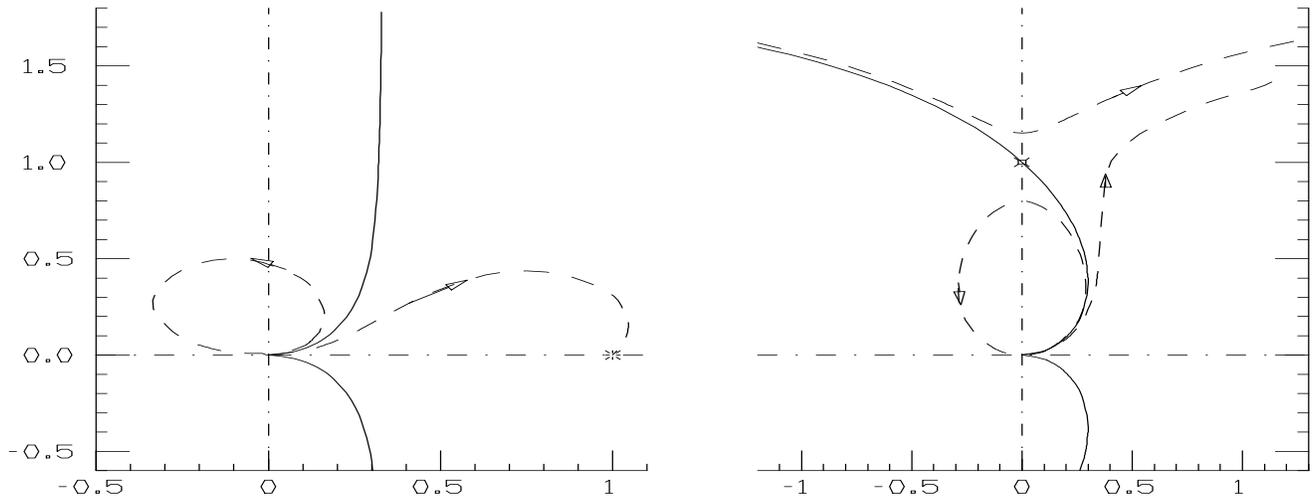}
\caption{Examples of RG trajectories (dashed) and separatrices (solid)
for the polynomial $\beta$-function (2.14a) with $\al_1\!=\!\infty$,
$\bal=1$ (left)
and for a singular $\be$ (2.14b) with $\gamma\!=\!0$, $h=1$ (right).}
\end{figure}

In Fig.1 examples of trajectories are shown together with the characteristic
lines -- separatrices -- which border PT and PH domains.
Crossing a separatrix in $\al$ causes non-analyticity in $I(\al)$.
It is worth reminding that the fact that the separatrices emerge
from the origin $+0$ along the circular {\em arcs}, is of the most general
nature and does not depend on the details of the model chosen for
illustration.
Thus, we arrive at the main conclusion of this Section,
that $I(\al)$ (for a rather trivial, purely perturbative, reason)
{\em cannot}\/ be analytic in any sector
with a finite opening angle $\theta_0$ covering the positive
real direction in the $\al$-plane.
As a result, the usual Borel representation of the physical quantity
$I^\PH(\al)$ does not exist.
Indeed, if the integral
$$
  \tilde I(\al) \>=\>
\int_0^{\infty} du\, B(u)\, \exp\left\{ -\frac{u}{\al}\right\}
$$
existed for some $\al=\al_0$,
it would define the function analytic for all
$\al$ with $\Re \al^{-1} > \al_0^{-1}$ which is the interior of
the circle $|\al-\half\al_0|=\half\al_0$.

This non-analyticity in $I(\al)$ at arbitrarily small positive $\al$
manifests itself in a singularity of the Borel image
on the real positive $u$-axis,
which is equivalent to the same-sign factorial asymptote
of the PT coefficients.
In the following we show that the Borel resummation actually yields
the {\em perturbative}\/ function (\ref{IPT}), $ \;\tilde I(\al)=I^\PT(\al)$.

\subsection{Borel transform for the PT phase}

Following the above line of reasoning, one concludes that the
analyticity domain in the {\em perturbative}\/ phase,
contrary to the physical phase, is broad enough
to allow one to represent $I^\PT(\al)$ in terms of the Borel integral.
Such a representation, however, involves {\em imaginary}\/ values of the
Borel parameter $u$.
Namely, for sufficiently small $\al$ in the upper half-plane ($\Im\al\!>\!0$)
one can write
\bminiG{PTBorel}
\label{Borel}
  I^\PT(\al) \>=\>
\int_0^{i\infty} du\, B(u)\, \exp\left\{ -\frac{u}{\al}\right\} \>.
\eeeq
The inverse relation gives the Borel transform $B(u)$ as a regular
function (for imaginary $u$),
\beeq
\label{InvBorel}
  B(u) \>=\> \int_{-\infty-i/2r}^{\infty-i/2r} \frac{dz}{2\pi i}\,
I^\PT(z^{-1}) \,  \exp\left\{ {u}{z}\right\}\>.
\emini
The integration line in $z=1/\al$ shown in (\ref{InvBorel}) corresponds
to a (small) circle in the upper half-plane of $\al$, namely,
$\abs{\al-ir}= r$.

In terms of the PT coefficients,
one substitutes the (asymptotic) PT expansion for $I$,
\beq\label{Iser}
  I^\PT \>=\> \sum_{n=1}^\infty a_n\, \al^n\>,
\eeq
into (\ref{InvBorel}).
Closing the contour by the (multiple) pole at $z\!=\!0$
($\Re uz \!<\!0$ for $\Im z>0$),
one obtains the standard improved series for the function $B(u)$
\beq\label{Bser}
 B(u) \>=\> \sum_{n=0}^\infty \frac{a_{n+1}}{n!}\, u^n\>.
\eeq
Thus, for {\em imaginary}\/ values of $u$ this series
determines the {\em regular}\/  Borel image $B(u)$.

\subsection{``Condensate'' contribution}
The IR renormalon problem arises when one attempts to reconstruct
the observable $I(\al)$ for {\em real}\/ $\al$ by means of the
perturbative series (\ref{Iser}) processed via the Borel machinery
(\ref{Bser}) and the integral representation (\ref{Borel}).
The latter now runs, however, along the {\em real}\/ positive $u$-axis.
Then $B(u)$ shows up a singularity (a pole, for the one-loop \bef, or a
cut in general)
at $u\!=\!u_0$
(in our normalization for the coupling, $u_0\!=\!p$).
As a result, the answer for $I$ becomes ambiguous and, generally,
complex, depending on the way one choses to pass by the singular point
on the integration line.

The PT coefficients $a_n$ do not care whether the expansion
parameter $\al$ has been chosen real or imaginary.
Therefore, we come to conclude that the IR renormalon is nothing but
the outcome of an unjustified attempt to force the perturbative answer
$I^\PT$ (\ref{IPT})
analytically continued outside its native phase, to represent the
true $I^\PH$.
Now we are in a position to cure the wound.
To do so we observe that the physical answer as given by (\ref{IPH})
can be reconstructed as a sum of two contributions:
the PT piece, $I^\PT(\al)$, analytically
continued to real $\al$ values {\em plus}\/ the ``condensate''
(or ``confinement'')
contribution originating from the discontinuity due to crossing
the separatrix on the way to the real axis.
Indeed, let us split the original integral along the physical
trajectory in (\ref{IPH}), running from $\bal$ to $\al$ into two pieces,
$-0\to\al$ and $\bal\to -0$, to write
\beq
 I^\PH(\al) = I^\PT(\al) \>+\> H(\al)\,,
\eeq
where
\beq\label{Hdef}
 H(\al) \>\equiv\>
  - \int^{\bal}_{-0} \frac{d\al'\>\al'}{\be(\al')}
\exp\left\{ -p\int^{\al}_{\al'}\frac{d\al''}{\be(\al'')}\right\}
\>=\> -f^p(\al)\int_{-0}^{\bal} \frac{d\al'}{\be(\al')}
\frac{\al'}{f^p(\al')}\,.
\eeq
The ``condensate'' contribution $H$ satisfies the homogeneous RG
equation (\ref{homo}).
Therefore, turning from the $\al$ representation to the momentum dependence,
one observes a pure power
\beq\label{Hexp}
 H(\al) \>=\> \cC\,f^p(\al)\>=\> \frac{{\cC}_\Lambda}{Q^{2p}}\,,
\eeq
with a {\em complex}\/ constant $\cC$
(or, equivalently, a dimensionful constant $\cC_\Lambda$)
to be determined from (\ref{Hdef}).
For example, for two models (\ref{twoex}) one has
\bminiG{2sols}\label{sol1}
 f(\al)= \al^{-\gamma}\,\exp\left\{ -\frac{1}{\al}\right\}
\left(1+\frac{\al}{\al_1}\right)^{\frac{\overline{\al}}{\al_1(\al_1+
\overline{\al})}}
\left(1-\frac{\al}{\overline{\al}}\right)^{-\frac{\al_1}{\overline{\al}
(\al_1+\overline{\al})}}
\eeeq
and
\beeq
 f(\al)=\al^{-\gamma}\,\exp\left\{ -\frac{1}{\al}+
\left(\frac{\gamma^2}{4}+h^2\right)
\al\right\}.
\emini
The ``condensate'' constant $\cC$ is expressed then, respectively,
in terms of the hypergeometric or the Bessel function, depending on $p$ and
the parameters of the \bef\refup{DUbig}.

{}From (\ref{Hexp}) the reason becomes clear why do we refer to $H$
as the condensate contribution (without quotation marks, from now on).
Let us draw the reader's attention to the fact that $I^\PH$ and $I^\PT$
have identical PT expansions, and $H$ has obviously none.

\mysection{Rescuing Borel representation: mission impossible}

A question arises, whether the Borel-type integral representation for the
physical answer can be rescued. At least within our simplified approach
in which the NP dynamics has been embodied into the behaviour
of the \bef\ at $\al\sim1$, the answer to this question is: in principle, yes.
However, as a more extensive analysis shows, the Borel construction one is
looking for proves to be not universal, depending essentially on the
NP features of the theory.
In other words, getting hold of the best possible PT information
appears to be insufficient for this purpose, calling for genuinely new,
non-perturbative, input.
The problem is mathematically more involved, so that in the present note
we give only a brief sketch of the procedure.

One starts by explicitly constructing the Borel image of $I^\PT$.
This can be done by translating the RG equation (\ref{RGeq}) into an
equation for $B(u)$.
To this end one writes (\ref{InvBorel}) in the form
\beq
   B(u) \>=\>\frac1{u-p} \int_C \;\frac{dz}{2\pi i}\:
\e\:^{pz}\,I^\PT(z^{-1}) \, \frac{d}{dz} \e\,^{(u-p)z}
\eeq
and, upon integrating by parts, makes use of (\ref{RGeq}) to arrive at
the following compact symbolic equation
\bminiG{Beqphi}\label{Beq}
uB(u)+ \phi\left(\frac1{\partial_u}\right)
\left\{\, (p-u)B(u)-1\,\right\} = 0\,; \quad
 \partial_u \equiv \frac{d}{du}\>.
\eeeq
The function $\phi$ here quantifies the deviation of the \bef\ from
its one-loop expression and is given by
\beeq\label{phidef}
 \phi(\al)\> =\>  \frac{\al^2}{\be(\al)+\al^2}\,.
\emini
For the case of the pure one-loop $\be(\al)\!=\!-\al^2$,
one has $1/\phi\equiv0$, and one gets a simple pole solution
\beq\label{Bone}
   B(u)=\frac{1}{p-u}\>=\> \frac1p\left(1-\frac{u}{u_0}\right)^{-1}\,,
\qquad u_0=p\>\left( = \frac{4\pi p}{\be_0} \quad
\mbox{in the standard normalization}\right).
\eeq
Given a (rational) \bef, it is straightforward to reduce (\ref{Beq})
to the differential equation (with proper initial conditions) for $B(u)$.
It is important to emphasize that the master (integral) equation
(\ref{Beqphi}),
as well as its differential counterpart,
is an equation with the only singular point
$$
  x\>\equiv\> 1-\frac{u}{u_0} \>\> =\>\> 0\,,
$$
which, in general, results in a branch point in the solution, $B(u)$,
at $u\!=\!u_0$.
Near the singularity
\beq\label{Bsing}
 B(u) = x^{-1}F(x) \>\approx\> c_0 \left(1-\frac{u}{u_0}\right)^{-1-p\gamma},
\eeq
where $\gamma$ is given by the first two loops of the \bef,
see (\ref{betaexp}).
Thus, one observes that both the position and the nature of
the Borel singularity have purely perturbative origin.
It is this universal singularity (\ref{Bsing}) that governs the same-sign
factorial growth of the PT series.

For example, within the models (\ref{twoex}) one arrives at a second order
differential equation.
In the first case (\ref{ex1}) it is a homogeneous confluent hypergeometric
equation for the function $F(x)=xB$,
with the initial conditions $F(1)=1$, $F'(1)=0$.
For the second model (\ref{ex2}) $B(u)$ is given by the solution of the
inhomogeneous Bessel equation (see \cite{DUbig} for details).

Having obtained $B(u)$, one then has to construct the Borel integral
\beq\label{rBorel}
 I^\PT(\al) = \int_0^\infty du\> B(u) \exp\left\{\frac{u}{\al}\right\},
\eeq
passing the singular point, say, from {\em above}\/ and to add the
condensate contribution due to crossing the {\em upper}\/ half-plane
separatrix.
{}From the Borel plane point of view, the latter contribution,
$H(\al)$, as a solution of the {\em homogeneous}\/ RG equation (\ref{homo}),
can be written (up to a factor) as an integral of the same function $B(u)$
along a {\em quasi-closed}\/ contour, namely,
around the cut $u_0<u<\infty$.
Thus, the full answer can be represented as a sum of two integrals,
one from zero to plus infinity (\ref{rBorel})
and the second embracing the cut (\ref{Bsing}).
The imaginary parts of these two contributions clearly cancel, as together they
constitute the physical answer $I^\PH$.

How much of the {\em real}\/ stuff remains?
To address this problem we restrict ourselves to
a polynomial \bef\ of power $n\!+\!2$.
In this case one has the $n$-th order differential equation to determine
$B(u)$ (generalized confluent hypergeometric equation),
with the initial conditions
$$
pB(0)=1\,,\qquad
\left.\left(\frac{d}{du}\right)^k\!\! (p-u)B(u)\right|_{u=0}\!\!=0\,,
\quad k=1\ldots n-1\,.
$$
Let us examine the {\em large}\/ $u$ asymptote of the solution.
For $u\!\to\!\infty$ equation (\ref{Beq}) reduces to
$$
0= \left[\, 1-\phi(\partial_u^{-1})\,\right] uB(u)
\;\; \Longrightarrow \;\;
\be(\partial_u^{-1}) uB(u)\,=0\,\;,
$$
which implies
\beq\label{polesum}
 uB(u) \simeq d_0\exp\left\{ \frac{u}{\bal} \right\}+
 \sum_i d_i\exp\left\{ \frac{u}{\bal_i} \right\} .
\eeq
Here $\bal_i\!\neq\!0$ is a set of zeroes of the \bef, other than the relevant
one, the physical fixed point $\bal$.
This particular term should cancel, however, in the combination of the
ordinary (PT) and the contour (NP) integrals (the discontinuity of $B(u)$ is a
solution of the homogeneous equation as well). Otherwise, the answer would be
singular at $\al=\bal$, which is not the case, since in the vicinity of
the fixed point it is analytic:
$$
 I^\PH(\al) = \frac{\bal}{p} \>+\> \cO{\al-\bal}\,.
$$
Moreover, for quite a while $I(\al)$ stays regular {\em above}\/ the fixed
point, $\al>\bal$, before the RG trajectories
betray $\bal$ for another, higher attractive point, i.e.
another (unphysical) NP phase occurs in which the coupling
is never small (no asymptotic freedom, that is).

{}From this consideration one concludes that if $\bal$ were the {\em only}\/
zero, the condensate contribution (contour integral) would have
to cancel the PT integrand above the singular point $u_0$
{\em completely}.
This is simply because there would be no other terms in the sum (\ref{polesum})
to maintain the asymptote.

This explains a miracle of the {\em finite}\/ Borel representation
one obtains within the model for the two-loop \bef\ with an
(anti-QCD) {\em negative}\/ $\gamma$.
One can get this model from (\ref{ex1}) setting $\al_1\!=\!\infty$
($\gamma=-\bal^{-1}$).
The master equation (\ref{Beqphi}) for the Borel image then becomes the first
order differential equation,
\bminiG{Gr}
  \phi(\partial_u)= \frac{1}{-\gamma}\partial_u = \bal\,\partial_u\;;
\qquad
\bal\frac{d}{du}\left[\,(p-u)B(u)\,\right] + uB(u)=0\;, \qquad B(0)=1/p\;.
\eeeq
Its solution,
\beeq\label{GBsol}
 B(u) = \frac{1}{p}\left(1-\frac{u}{p}\right)^{-1-p\gamma}
\!\!\exp\left\{\frac{u}{\bal}\right\}
\>\equiv\> \frac{1}{p}\left(1-\frac{u}{u_0}\right)^{-1+{p}/{\bal}}
\!\!\exp\left\{\frac{u}{\bal}\right\},
\emini
reveals explicitly the general features, namely, the nature of
the singularity at $u_0\!=\!p$ and the expected exponential behaviour
at infinity.

To evaluate the condensate contribution one first puts
$1/\al_1\!=\!0$ in (\ref{sol1}) to fix
$$
 f(\al)= \left(\frac{\bal-\al}{\al}\right)^{-1/{\bal}}
\exp\left\{-\frac{1}{\al}\right\} \,.
$$
Then, one performs integration in (\ref{Hdef}) to obtain
\beq\label{Gcond}
\eqalign{
\cC &= -\int_{-0}^{\bal} \frac{d\al'}{\be(\al')} \frac{\al'}{f^p(\al')}
=  \int_{-0}^{\bal}\frac{d\al'}{\al'(1-\al'/\bal)}
\left(\frac{\bal-\al'}{\al'}\right)^{\gamma_p}
\!\!\exp\left\{\frac{p}{\al'}\right\} \cr
&= -\int_{-\infty-i0}^1 dz\, (z-1)^{-1+\gamma_p} \exp\{\gamma_p z\}
\>=\>\left(\frac{\e}{\gamma_p}\right)^{\gamma_p}\! \Gamma(\gamma_p)
\, \e^{-i\pi\gamma_p}\>; \qquad \gamma_p=-p\gamma=\frac{p}{\bal}>0\,.
}\eeq
Finally,
\beq
 H(\al)= \left(\frac{\e}{\gamma_p}\right)^{\gamma_p}\! \Gamma(\gamma_p)
\, \e^{-i\pi\gamma_p} \cdot \left(\frac{\al}{\bal-\al}\right)^{\gamma_p}
\exp\left\{-\frac{p}{\al}\right\}\,.
\eeq
This expression, upon inspection, differs only by sign from the part of the
ordinary (PT) Borel integral from $u_0$ to $\infty$,
$$
 H(\al) = -\int_{u_0}^{\infty+i0} du\, B(u)\,
\exp\left\{ -\frac{u}{\al}\right\}.
$$
As a result, for the physical answer in this model the {\em finite-support}\/
Borel-type representation holds,
\beq
 I^\PH(\al) = \int^{u_0}_0 du\, B(u)\, \exp\left\{ -\frac{u}{\al}\right\}.
\eeq
This observation has been independently made in \cite{Grun}.

We now try to examine on this explicit example the standard routine employed in
the renormalon analysis. Namely, it is accustomed to take the PV of the
standard Borel integrals which amounts, for this model,
to taking the value
\beq
I_B(\alpha)\;\equiv \; I^\PH(\alpha) -\Re H(\al)\;=\; I^\PH(\alpha) -
f^p(\al) \left(\frac{\e}{\ga_p}\right)^{\ga_p} \Gamma(\ga_p) \cos{\pi\ga_p }\,.
\label{69}
\eeq
At the same time, the uncertainty associated with the Borel resummation
procedure is usually estimated as
\beq
\delta I(\alpha)\;=\; \frac{1}{2\pi}\;\abs{
\int_{\cal C} \; du\: B(u)\, \e^{-u/\al} }\;,
\label{70}
\eeq
with the contour running around the cut, which in our case yields
\beq
\delta I(\alpha)\;=\;f^p(\al) \left(\frac{\e}{\ga_p}\right)^{\ga_p}
\Gamma(\ga_p)\, \frac{1}{\pi}\abs{\sin{\pi\ga_p }}\,.
\label{71a}
\eeq
Both the error of the PV Borel summation
$\Delta_B=I^\PH\!-\!I_B$ and the
estimated uncertainty $\delta I$ have the same correct power behavior
$(\Lambda^2/Q^2)^p$.
However, their relative magnitude, in general, mismatches:
\beq
\frac{\Delta_B}{\delta I}=\frac{\pi}{\abs{\tan{\pi\ga_p}}}\,,
\label{71}
\eeq
so that the actual error can become much larger than the estimated one, which
happens, for example, when $\ga_p$ is numerically
small\footnote{We parenthetically note that it is actually the case
for the limit $\be_0 \to \infty$ (or $n_f \to -\infty$),
 viz., one gets $\ga \sim 1/n_f$
if assumes that the higher order terms in the
\bef\ can be in turn obtained by only the leading in $ 1/n_f$ contributions,
though literally one gets positive $\gamma$.}.
The origin of such a mismatch is readily understood\refup{DUbig}.

It is worthwhile to mention another interesting feature of an interplay between
PT and NP contributions.
Namely, when $\ga_p$ happens to be a positive integer,
the singularity (\ref{Bsing})
disappears and $B(u)$ remains {\em analytic}\/ in the entire complex
$u$-plane.
There is no ambiguity in chosing the path of integration,
and the Borel summation yields unambiguous -- but incorrect --  result!
What is more intriguing from the viewpoint
of the naive PT analysis is that the factorial growth of
the coefficients in the expansion of $I(\al)$ disappears:
\beq
a_n\;\sim\; n^{\ga_p-1}\,(\bal)^{-n}\, .
\label{73}
\eeq
Resummation of the ``subleading'' $1/n$ corrections to the asymptote of $a_n$
kills factorials altogether.
Analyzing mere perturbative series one would not infer any deficiency
of the expansion, although the presence of the ``NP condensate''
has been demonstrated explicitly.

For illustration, in the simplest case $\ga_p\!=\!1$ one has
\beq
I(\al)\>=\> \frac{\al\,\bal}{p(\bal-\al)}
\left(1\,-\, \exp\left\{1-\frac{\bal}{\al}\right\}\right).
\eeq
The first term is what one gets by the PT summation whereas the second
one is the NP contribution.
The former is merely a geometric series in $\al$, showing no indication
of NP effects.
The latter term, on the contrary, has literally no PT expansion.
In general, however, the situation is different and such an apparent
splitting is absent, so that PT and NP pieces cannot be explicitly
separated.
It is just the genuine case of intrinsically mixed PT and NP contributions
when the PT series normally send the message, via IR renormalons,
about the presence of the non-perturbative effects.
Notice, that for a generic \bef ~($n\!>\!1$) an integer $\ga$ does not
mean analyticity: $u_0$ becomes a logarithmic branch point of $B(u)$.

A curious observation is that for a given (polynomial) \bef\
one can construct a polynomial $P_n(\al(k^2))$
to replace $\al(k^2)$ in the definition of the ``observable''
(\ref{observ}), such that $B(u)$ becomes analytic and, thus, the  PT expansion
factorial-free\refup{DUbig}.
However, such a construction proves to be non-universal with respect to $p$.

In conclusion, let us mention that an attractive scenario of the finite-support
Borel representation yielding $I^\PH(\al)$ analytic everywhere but at $\al=0$,
is a peculiar property of the oversimplified two-loop model.
Elsewhere such a possibility may accidentally occur only at specific,
contrived values of the parameters.
However, this would hold, once again, for one particular value of $p$,
while for observables with the canonical dimension other than $2p$ the
contribution from $u\!>\!u_0$ gets resurrected.
Therefore we consider such an accident carrying no physical significance.

\mysection{Toy OPE for the Toy model}

For short distance Euclidean observables the Wilson OPE is known to provide
a proper framework for describing power suppressed effects.
It automatically cures, once and forever, the IR renormalon trouble
for the price of modifying perturbative coefficients \cite{fail}.
The latter are changed only a little in low orders but become completely
different at $n\!\to\!\infty$. How does it happen within our toy model?

The analog of the OPE for $I(\al)$ is introducing a normalization point
$\mu^2\ll Q^2$ (but still belonging to the small coupling domain)
and considering
\beq
I(\al)\;= \; I_\mu^{\rm ld}(\al)\;+ \; I_\mu^{\rm sd}(\al)\;\equiv \;
\int_0^{\mu^2}\; \frac{dk^2}{k^2}
\left(\frac{k^2}{Q^2}\right)^p \,\al(k^2)\;+
\int_{\mu^2}^{Q^2}\; \frac{dk^2}{k^2}
\left(\frac{k^2}{Q^2}\right)^p \,\al(k^2)
\;\; .
\label{ope}
\eeq
Now one treats perturbatively only the $I^{\rm sd}(\al)$ part.
Following the above analysis, we want to study analytic properties
of $I^{\rm sd}(\al)$ at $\al\!=\!0$.
It is important that $\mu$ is kept fixed so that $\al(\mu^2)$
depends on $\al$ via
\beq
\int_\al^{\al(\mu^2)}
\;\frac{d\al'}{-\be(\al')}\;=\;\log{\frac{Q^2}{\mu^2}} = \tau\, .
\label{tau}
\eeq
It implies that the RG trajectories employed for calculating
$I_\mu^{\rm sd}(\al)$ are always of the fixed finite time interval
$\tau\!=\!\log{Q^2/\mu^2}$.
This is what makes the crucial difference:
$I^{\rm sd}(\al)$ is an {\em analytic}\/ function at $\al\!=\!0$
with a finite radius of convergence!

In reality, one is interested in a proper value of $\mu$ necessary for
the OPE which, for practical reasons, is desired to be as low as
possible.
One should bear in mind that now the coefficients of the perturbative series
for $I_\mu^{\rm sd}$ depend explicitly on the ratio $\mu^2/Q^2$.
Under such circumstances a critical momentum scale $\mu_{\rm sh}$
emerges \cite{DUbig} such that the series converge for
$\al\!<\!\al(\mu^2)$
if $\mu$ is chosen above $\mu_{\rm sh}$,  and diverge
(for sufficiently small $\al$, that is, large $\ln(Q^2/\mu^2), sic!$)
if $\mu\!<\! \mu_{\rm sh}$.
We denote the value of the coupling corresponding to this scale by
$\al_{\rm sh}= \al(\mu_{\rm sh})< \bal$.
Leaving aside tiny details concerning perverted \bef{s} \cite{DUbig},
one can determine this OPE-borderline value solving the equation
\beq
\Re \int_{\al_{\rm sh}}^{\infty} \;\frac{d\al'}{-\be(\al')}\;=\; 0\,,
\label{alc}
\eeq
where the integration contour travels to infinity along a separatrix.
Such an equation always has a unique solution between zero and $\bal$.
In a general case when the structure of zeroes is rich and there is a set
of separatrices, one should take the minimal solution.
(Singularities of the \bef, if any, should also be considered,
with the position of the singularity replacing $\infty$
in the integral in (\ref{alc})).
For example, in the two-loop model (\ref{Gr})
one obtains $\al_{\rm sh}\simeq 0.782\,\bal$.
For the model (\ref{ex2}) with $\gamma\!=\!0$, discussed in \cite{KP},
$\al_{\rm sh}=h^{-1}$.

Borel non-summability in $I$ within the OPE lies in the
second, large distance, piece $ I_\mu^{\rm ld}$, which, of course,
cannot be even dreamed to be calculated in terms of $\al$ expansion.
This contribution is explicitly associated
with the particular, physical, phase.

We note that an attempt to do the OPE without an IR cutoff, i.e. putting
$\mu\!\to\! 0$ (or, equivalently, using  Dimensional Regularization to
``renormalize'' the power convergent integrals like $I$)
means calculating $I^\PT(\al)$
instead of $I_\mu^{\rm sd}(\al)$, and leaving only $H(\al)$ in
$I_\mu^{\rm ld}(\al)$.
In other words, it is an attempt to entirely subtract
``perturbative corrections'' from $I^{\rm ld}$.
It is just this action that generates the IR renormalons
in the coefficient functions making them uncalculable.
As long as the exact analytic expression for  $I(\al)$ is
known, such a routine seemingly does not pose particular problems.
However, in practical applications it looks rather dangerous
when multiloop corrections are incorporated.
Indeed, consider for example the $Q^2$-dependence of the observable.
Translating the behaviour in $\al(Q^2)$ into the $Q^2$-dependence
one finds a smooth behaviour of $I$ at small $Q^2$:
\beq
I(\al(Q^2))
\>\simeq\> \frac{\bal}{p} \>+\> \frac{1}{p+\rho}
\left(\frac{Q^2}{\L^2}\right)^\rho, \quad \rho\equiv
\left.\frac{d\be(\al)}{d\al}\right|_{\al=\bal}>0\,;
\qquad \mbox{for}\>\>  Q<\L\>.
\label{94}
\eeq
On the contrary, according to (\ref{Hexp}),
both the ``perturbative''  and ``purely non-perturbative'' parts,
taken separately, are singular:
\beq
 I^\PT(\al(Q^2)) \>\sim\> H(\al(Q^2)) \>\propto\>
\left(\frac{\L^2}{Q^2}\right)^{p} .
\label{95}
\eeq
Therefore, the naive Borel resummation
typically yields a strongly corrupted picture of the actual scale
of long distance phenomena.
We will return to this important point in \cite{DUbig}.

\mysection{Conclusions}

We argued that the IR renormalons have a transparent mathematical origin
and are associated with employing the same short distance coupling
as an expansion
parameter for describing physics at both large and small scales.
They do not actually show what happens at low momentum scales but only
send a message that something {\em may} happen there.
Inspired by this idea\refup{ackn} we studied a simplified model for hard
QCD observables in which all the details of strong dynamics are embodied
into an IR non-trivial \bef.

We explicitly showed that, contrary to a rather popular believe,
IR renormalons are not directly related to the Landau singularity in the
running coupling.
Even if it freezes at a small value, the same-sign
factorial growth of the PT coefficients remains intact.
At the same time, there are curious examples when no IR renormalon is
around (the PT series converges!) but the PT expansion is as deficient
as in a general case.

The position and the nature of the IR renormalon singularities
are determined by the first two coefficients of the \bef, i.e.
are governed by the deep PT domain.
The true origin of the IR renormalons is rooted in a non-trivial phase
structure of QCD.

We examined how well can renormalons represent, in general, the actual
long distance effects and found them unsatisfactory in many important
respects including the absolute magnitude, an estimate of incompleteness of the
purely PT approximation and the steepness of the $Q^2$-dependence in the
transition regime from short distances to the strong interaction domain.

As far as the {\em ultraviolet}\/ renormalons are concerned,
in the framework of our model they appear in the observables
represented by an integral
\beq\label{uv}
 U(\al) = \int_{Q^2}^{\infty} \frac{dk^2}{k^2} \left(\frac{Q^2}{k^2}\right)^p
\al(k^2)\>.
\eeq
Straightforward application of the analysis outlined in the present paper
shows that $U$ is analytic at small $\al$ except for only a narrow beak
around the real {\em negative}\/ axis.
Therefore, its Borel resummation yields the correct result.
This being a positive statement, we are not sure of how heavily does
the conclusion rely on particular features of the oversimplified model.
Therefore we refrain from definite claims of its applicability to actual QCD.

\vspace{2 mm}

When this paper was in writing we learned about
the preprint by G.~Grunberg \cite{Grun} where the finite support
Borel representation has been constructed for the two-loop (``non-QCD'')
\bef,  in agreement with our finding.
\vspace*{0.5cm}

\noindent
{\bf Acknowledgements:}\\
N.U. gratefully acknowledges crucial insights from A. Vainshtein and
an important impetus of collaboration
with M.~Shifman. He is also indebted to I.~Bigi
and L.~McLerran for the encouraging interest.
We thank V.~Braun and  J.~Fischer for illuminating discussions.
This investigation was promoted by the informal atmosphere of the CERN
TH Renormalon Club and we thank its participants for interest.
\vspace*{3 mm}

\noindent{\large\bf References}
\begin{enumerate}
\item\label{recent}
Listing of even the most relevant papers would be too extensive,
and we leave it for the forthcoming detailed presentation \cite{DUbig}.

\item\label{ren}
G. 't Hooft, {\bf in} {\em The Whys Of Subnuclear Physics}, Erice
1977,
ed. A. Zichichi (Plenum, New York, 1977);
B. Lautrup, {\it Phys. Lett.} {\bf 69B} (1977) 109;
G. Parisi, {\it Phys. Lett.} {\bf 76B} (1978) 65; {\it Nucl. Phys.}
{\bf B150} (1979) 163;
A. Mueller, {\it Nucl. Phys.} {\bf B250} (1985) 327.\\
For a recent review see\\
A. H. Mueller, {\bf in} {\em Proc. Int. Conf. ``QCD -- 20 Years Later"},
Aachen 1992, eds. P. Zerwas and H. Kastrup, (World Scientific,
Singapore, 1993), vol. 1, page 162.

\item\label{DUbig}
Yu.L. Dokshitzer, N.G. Uraltsev, {\it in preparation}.

\item\label{blm}
 S.J. Brodsky, G.P. Lepage and  P.B. Mackenzie, {\it Phys. Rev.} {\bf D28}
(1983) 228;\\  G.P. Lepage and  P.B. Mackenzie, {\it Phys. Rev.} {\bf D48}
(1993) 2250.

\item\label{Grun}
G. Grunberg, {\bf preprint} CNRS UPRA 0014  [hep-ph/9511435].

\item\label{fail}
V. Novikov, M. Shifman, A. Vainshtein and V. Zakharov, {\it Nucl. Phys.}
{\bf B249} (1985) 445;\\
In the context of the heavy quark expansion these ideas were reiterated
recently in:\\ I.I. Bigi, M. Shifman, N.G. Uraltsev, A. Vainshtein,
{\it Phys. Rev.}  {\bf D50} (1994) 2234.

\item\label{KP}
N.V. Krasnikov, A.A. Pivovarov, {\bf preprint} INR 0903/95 [hep-ph/9510207].

\item\label{OPE}
For a review see: M. Shifman, Ed., {\em Vacuum Structure and QCD Sum Rules},
North-Holland, 1992, Chap.~3.

\item\label{ackn}
Our view upon IR renormalons as a short distance effect
followed from the discussion of N.U. with A.~Vainshtein a year ago
and reportedly ascends to V.I.~Zakharov.

\end{enumerate}

\vfill
\end{document}